\input{aipcheck.tex}

\documentclass[
 4apaper
  ]
  {aipproc}

\layoutstyle{8x11single}

\SetInternalRegister\hbadness{8000}

\begin{document}

\title {Spatial features of non-thermal SZ effect in galaxy clusters}


\author{E. Palladino}{
  address={Università "La Sapienza", P.le Aldo Moro 5, Rome,
  Italy},
  email={emilia.palladino@roma1.infn.it}
}

\author{S. Colafrancesco}{
  address={Osservatorio Astronomico di Roma, Monteporzio, Italy},
  email={cola@coma.mporzio.astro.it}
}

\author{P. Marchegiani}{
  address={Osservatorio Astronomico di Roma, Monteporzio, Italy}
}

\copyrightyear  {2001}

\begin{abstract}
We investigate the spatial behaviour of the total comptonization
parameter $y_{tot}$ evaluated for a galaxy cluster containing two
population of electrons: the thermal population, with energy
around some KeV and whose trace is evident in the X-ray emission
of the ICM (Intra-Cluster Medium), and the relativistic
population,  which give rise to the radio halo emission found in
several clusters of galaxies. We present the first results
obtained from our analysis showing that there are remarkable
features in such spatial trend, which might throw a new light in
understanding the cluster internal processes.
\end{abstract}

\maketitle

Recently has been investigated the possibility that the thermal SZ
effect, {\em i.e.} the shift in  photon energy of the cosmic
microwave background radiation (CMBR) due to its passage through
the intra-cluster medium (ICM) (\cite{sz72}, \cite{rep95},
\cite{birk99}), could be implemented by the so called non-thermal
effect due to a relativistic electron population with energy of
order tens of KeV, instead of the usual thermal amount of some KeV
(\cite{ek2000}, \cite{bso2000}).

Here we stress the fact that the presence of a non-thermal SZ
effect also produces spatial variations of the total SZ radial
profile of the cluster.

Let us consider a galaxy cluster in which are present two kind of
electron populations: the first one is {\em thermal}, {\em i.e.}
it follows a relativistic maxwellian velocity distribution, the
second one is {\em non thermal}, {\em i.e.} its energy spectrum is
relativistic in a certain range. In particular we have used a
phenomenological complex spectrum (double-power law) with slopes
$\alpha_r \sim 2.5 $ at $E > E_*$ and $\alpha_x \sim -0.5 $ at $E
< E_*$ with the break set at $E_* \sim 200 \div 400$ MeV, suitable
to reproduce the whole set of non thermal phenomena found in
several galaxy clusters \cite{petrosian} . Consequently, the
behaviour of this second population can not be treated with the
Kompaneets approach usually used to describe the dynamic of the
thermal one (for further details on a complete relativistic
treatment of the non-thermal effect see \cite{noi}) . In a first
approximation we consider these populations independently
superposed, so it is possible to completely separate their
contribution to the total SZ effect on the incoming photons of the
CMBR (Colafrancesco, Marchegiani \& Palladino, these Proceedings,
for more details).

The inverse Compton scattering of the CMB photons by the electrons
of the ICM is theoretically described by the so-called {\em
comptonization parameter}. For a thermal radiation field and in
the more general form such a contribution can be expressed as
$
y_{th}=y_{0,th}\cdot g(x),  
$
where $y_{0,th}$ physically represents the time spent by the
radiation in the medium and it is proportional to the integral of
the thermal electron kinetic pressure $P_{th}= n_e k_B T_e$, while
the function $g(x)$ contains the dependence of the effect from the
a-dimensional frequency $x=h\nu /k_{B}\,T_{CMB}$ of the background
radiation.

According to the classical approach we can describe the kinetic
pressure $P_{th}$ using a theoretical model in which we assume
known the spatial distribution and the temperature of the electron
population. The model we have used is the {\em isothermal $\beta
$-model} \cite{bmodel},
according to which the temperature of the ICM is constant and equal to $%
T_{e} $ and the electron density $n_e$ has a spherical
distribution parameterized by the relation
\begin{equation}
n_{e}({\bf r})=n_{e0}\left[ \,1+\left( \frac{r}{r_{c}}\right) ^{2}%
\right] ^{-3\beta /2} , \label{betamodel}
\end{equation}
where $n_{e0}$ is the electron density at the center of the cluster, $%
r_{c}$ is the core-radius and $\beta = \mu m_p v^2 /k_B T_e$ is
the exponent observed to be in the range 0.6 $-$ 1 (see, e.g.,
\cite{sar88}). The detailed values of such parameters are found
from the analysis of X-ray emission maps produced by the thermal
electrons of the cluster.

Using the model \eqref{betamodel}, calling
$P_{th}^{0}=n_{e0}k_{B}T_{e}$ the central kinetic pressure and
calculating the integral along the line of sight \cite{sar88}, we
obtain the final expression for the thermal comptonization
parameter:
\begin{equation}
y_{0,th}=\frac{\sigma _{T}}{m_{e}c^{2}}\,P_{th}^{0}\,r_{c}\,Y_{th}(%
\theta, \theta_c ).  \label{ythfin1}
\end{equation}
In the function $Y_{th}$, whose explicit expression is given by
\begin{equation}
Y_{th}(\theta, \theta_c )=\sqrt{\pi }\;\frac{\Gamma (\frac{3}{2}\beta -\frac{1}{2})%
}{\Gamma (\frac{3}{2}\beta)}\,\left[ 1+\left( \frac{\theta
}{\theta _{c}}\right) ^{2}\right] ^{\frac{1}{2}-\frac{3}{2}\beta},
\label{ythspace}
\end{equation}
is contained the dependence of the effect from the spatial
coordinates, {\em i.e.} from the angle $\theta$ between the center
of the cluster and the direction of observation, while
$\theta_{c}=r_{c}/D_{A}$ is the angular core radius as deduced
from the X-ray data. The {\em angular diameter distance}, $D_{A}$,
of the cluster, given in terms of the redshift, contains all the
information relative to the assumed cosmological model; in all our
work we have put null the cosmological constant and the curvature
and the Hubble constant is in the form $H_{0}=h_{50}50$ km
sec$^{-1}$ Mpc$^{-1}$, with $h_{50} =1$.

Taking into account that we use the same formalism to describe
both the thermal and non thermal contributions to the total
effect, we can soon write down all the expressions we have
previously obtained also in the case of the relativistic
population, in which we again parameterized the electron density
with a spherical distribution, {\em i.e.} with a $\beta$-model.
One fundamental aspect to underline is the difference between the
relativistic distribution and the thermal one, which consists in
the fact that the spherical radius of the first distribution must
be truncated at some limiting radius $r_{lim}$, as it results from
the measurements of the radio emission from clusters of galaxies,
that in the case for example of COMA is found to be
$r_{lim}=1.25\,h_{50}^{-1}$ Mpc.

So for the non thermal comptonization parameter we have:
\begin{equation}
y_{0,non-th}=\frac{\sigma_{T}}{m_{e}c^{2}}\,P_{th}^{0}\,r_{c,rad}\,\bar{P}_0%
\,Y_{non-th}(\theta, \theta_{c,rad} ),  \label{ynonthfin1}
\end{equation}
where the subscript ``$rad"$ is used to remember that these
quantities refer to the relativistic electrons that emits in the
radio band of the electromagnetic spectrum (they in general belong
to the {\em radio halo} of the cluster), while $\bar{P}_0
=P_{rel}^{0}/P_{th}^{0}$ is the ratio between the relativistic and
the kinetic pressures of the electrons in the centre of the
cluster. The function $Y_{non-th}(\theta, \theta_{c,rad} )$, in
which $\theta _{c,rad}=r_{c,rad}/D_{A}$ is the angular core radius
of the radio halo, has the same expression as the one of eq.
\eqref{ythspace}, in which we substitute the parameters proper of
the non-thermal effect.

Now we have all the necessary information to put together the two
previous contributions in a whole expression for the total
comptonization parameter $y_{tot}$ given by the sum
$y_{th}+y_{non-th}$. Elaborating this expression we have the final
relation used to obtain our results (here the subscript ``$X"$
refers to the thermal quantities given by the clusters $X$-ray
maps):
\begin{equation}
y_{tot}=\frac{\sigma_{T}}{m_{e}c^{2}}\,P_{th}^{0}\,\left\{
r_{c,X}\,Y_{th}(\theta, \theta_{c,X} )\cdot
g(x)+r_{c,rad}\,\bar{P}_0\,Y_{non-th}(\theta,\theta_{c,rad} )\cdot
\tilde{g}(x)\right\} .  \label{ytotfin}
\end{equation}

In Tab.\ref{tabparam} we present the parameters used to make the
analysis and to obtain the results we present for the COMA
cluster.

\begin{table}
\begin{tabular}{rrrrrrrrr}
\hline
   \tablehead{1}{r}{b}{$z$\\}
  & \tablehead{1}{r}{b}{$k_B T_e$\\ ($keV$)}
  & \tablehead{1}{r}{b}{$r_{c,X}$\\ ($h_{50}^{-1} Mpc$)}
  & \tablehead{1}{r}{b}{$r_{c,rad}$\\ ($h_{50}^{-1} Mpc$)}
  & \tablehead{1}{r}{b}{$r_{lim,X}$\\ ($h_{50}^{-1} Mpc$)}
  & \tablehead{1}{r}{b}{$r_{lim ,rad}$\\ ($h_{50}^{-1} Mpc$)}
  & \tablehead{1}{r}{b}{$\beta _{X}$\\}
  & \tablehead{1}{r}{b}{$\beta _{rad}$\\}
  & \tablehead{1}{r}{b}{$n_{e0,X}$\\ ($h_{50}^{2}\; cm ^{-3}$)} \\
\hline 0.0232 & 8.5 & 0.42 & 0.4 & 4.2 & 1.25 & 0.75 & 0.8 & 3.0
$\cdot 10^{-3}$
\\ \hline
\end{tabular}
\caption{Values of the parameters used for the COMA cluster study}
\label{tabparam}
\end{table}

\begin{table}
\begin{tabular}{cccc}

\hline
   \tablehead{1}{r}{b}{Channel}
 &  \tablehead{1}{r}{b}{$\lambda$ (mm)}
 &  \tablehead{1}{r}{b}{$\nu $ (GHz)}
 &  \tablehead{1}{r}{b}{$x$}   \\
\hline 1 & 2.1 & 142.857 & 2.51 \\ 2 & 1.4 & 214.286 & 3.77\\ 3 &
1.1 & 272.727 & 4.80\\ 4 & 0.85 & 352.941 & 6.21\\ \hline
\end{tabular}
\caption{The four values of $x$ chosen to make the analysis (see
text for details).} \label{tabx}
\end{table}

In Fig.1 we show the behaviour of the comptonization parameter,
where we have fixed four values of the adimensional frequency $x$
(see Tab.\ref{tabx}), to simulate the frequency channels in a
common observational experiment devoted to measure clusters
properties at millimeter and sub-millimeter wavelength and three
values of the ratio $\bar{P}_0$ between the thermal and
relativistic pressures of the electron populations, {\em i.e.}
0.05, 0.49, 1.48. In each panel we graph the spatial trend of the
total SZ effect comptonization parameter $y_{tot}$ together with
the thermal one $y_{th}$ founded for the same physical values, to
make a comparison among the different curves, from the minimum of
the effect ($x_{min} \sim 2.5$), the zero ($x_0 \sim 3.8$), to its
maximum ($x_{max} \sim 6.2$).

As one can see, the detection of non-thermal effect is greater in
the first two channels, even if the second channel presents a
difficult in measuring the effect for the presence of the maximum
of the kinetic one that can be in principle greater then the
non-thermal one. Going towards higher frequencies ($x > 5$), we
find that there is no observational evidence of the non-thermal
effect with respect to the thermal one.

Another point is the jump one can see in both the upper panels: it
is due to the fact that the relativistic population has a core
radius smaller than the optical radius of the cluster (that is
given by $r_{c,X}$). The possibility to detect such a jump in the
spatial distribution is linked to the angular resolution of the
experiment devoted to measure the effect: it is possible, in fact,
that a window function of many arc-min, with respect to the
angular dimension of the examined cluster, cannot see such a
characteristic spatial feature, for the signal has to be
convoluted with the instrumental response.

In conclusion we underline the importance of specific spectral and
spatial features of the non-thermal SZ effect that can be detected
through a multi-frequency observation with narrow-band detectors:
the best observational strategy is to check the frequency range $x
\sim 2 \div 8$ where the spectral features allow to distinguish
the non-thermal effect to the thermal one.

This is very important since the SZ effect is a remarkable tool
for cosmology ({\em i.e.} with the measurement of the Hubble
constant $H_0$) and for better knowing the astrophysics of
clusters. We stress that the PLANK surveyor experiment has the
possibility to detect the total SZ effect (thermal and non-thermal
one) in a a large number of nearby radio-halo (with a relativistic
electron component) clusters.

\begin{figure}
\includegraphics{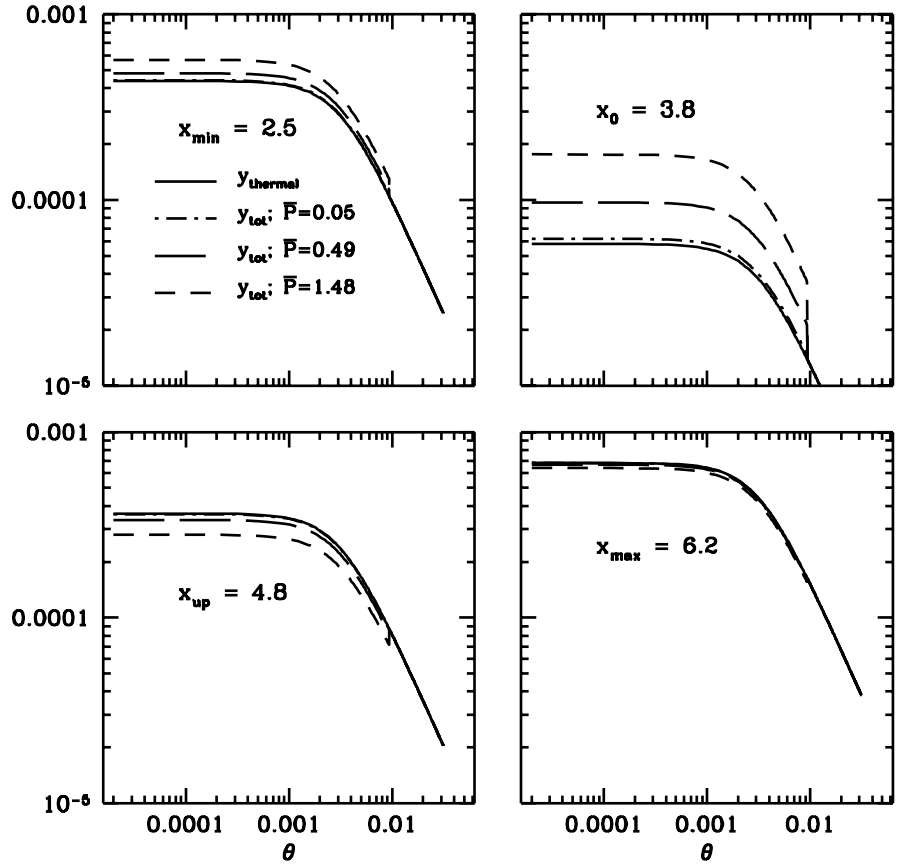}
\caption{The spatial dependence of the total SZ effect produced by
the combination of the thermal and relativistic electrons in COMA,
with respect to the thermal one ({\em thin line}). The four panels
refers to the four values of the frequency $x$ and in each one
there are plotted the case for $\bar{P}_0 =0.05$ ({\em point-dash
line}), for $\bar{P}_0 =0.49$ ({\em long-dash line}) and
$\bar{P}_0 =1.48$ ({\em short-dash line}).}
\end{figure}

\end{document}